\documentclass[%
 reprint,
 amsmath,amssymb,
 aps,
]{revtex4-2}

\usepackage{graphicx}
\usepackage{dcolumn}
\usepackage{bm}

\begin{document}


\title{Characteristic transition of the dominant power loss from diffractive to ohmic \\in overmoded and periodically loaded  waveguides at high frequency}

\author{Adham~Naji$^1$}\email{anaji@scu.edu}
 \altaffiliation[]{}
\author{Pawan~Kumar~Gupta$^1$}
\author{Gennady~Stupakov$^2$}
 
\affiliation{%
 $^1$Santa Clara University, Santa Clara, California, 95053\\
 $^2$SLAC National Accelerator Laboratory, Stanford, California, 94025
}%

\date{\today}

\begin{abstract}
The analysis of electromagnetic fields in cylindrical waveguiding structures that contain periodic ring loading, whether for applications in charged-particle accelerators or radiation transportation, has been traditionally conducted under simplifying limits, where the structure is either single-moded at the lower-frequency limit or overmoded at high-frequency limit. These limits have often allowed us to find spectral (modal) expansions for the fields under simpler analytical and computational conditions, with ohmic effects typically being the dominant power loss mechanism in the lower limit, while diffraction effects dominate the loss in the higher limit. In this Letter, we report the observation of a transition point in the character of the main loss mechanism, where ohmic loss becomes dominant in a structure typically presumed to be dominated by diffraction loss. The results follow a formal analysis for the scattered vector fields in a highly overmoded THz waveguide. The findings bridge between the traditional theoretical descriptions for the two limits and reveal key tradeoffs that inform experiments for the transportation of THz radiation over long distances.
\end{abstract}

\maketitle



\begin{figure*}
\includegraphics[width=0.85\textwidth]{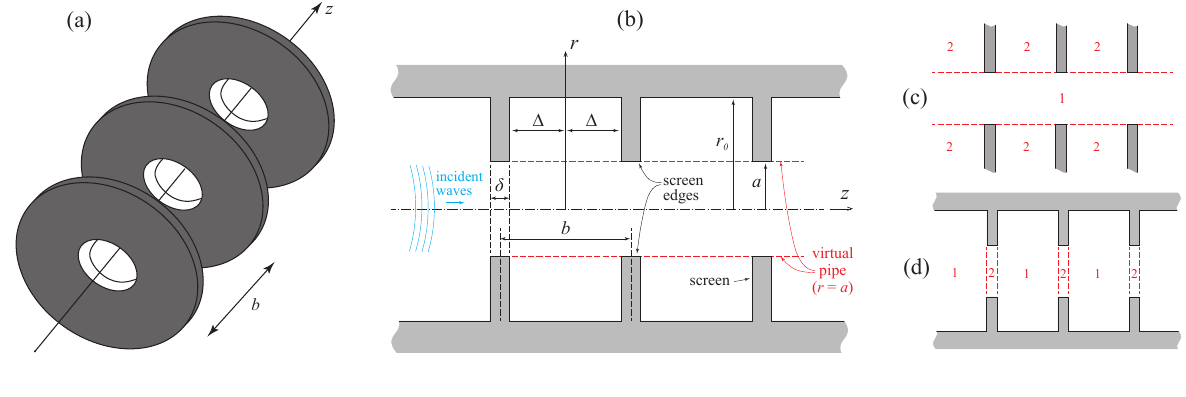}
\caption{\label{fig:geometry}  (a) Geometry of the structure in three-dimensions, showing the screens without the enclosing chamber at the outer radius. (b) A cross section, showing the enclosing chamber. (c) Division of space into region 1 (axial) and 2 (annual) for eigen-field analysis \cite{NajiTHz1}. (d) Division of space into region 1 (cavity) and region 2 (waveguide) for scattered-field analysis \cite{NajiTHz2,NajiTHz3}.}
\end{figure*}

\emph{Introduction}---Classical cylindrical waveguides that support electromagnetic wave propagation have been studied for decades and used in different forms and configurations \cite{jackson,Collin2}. They are indispensable to many applications that rely on the transportation of, or the interaction with, radiofrequency (rf), microwave, or THz electromagnetic fields. Examples include charged-particle accelerators, high-power light sources, radar, medical therapy instrumentation, and telecommunication systems.  
The periodically loaded cylindrical structure, whose general geometry is similar to the one shown in Figure~\ref{fig:geometry}, is one of the most common and versatile wave guiding structures in rf physics \cite{jackson,Collin2,Chao,slaterbook,schwinger,Mahmoud}. The periodicity of the geometry permits Bloch-like waves to propagate, with $k$-$\omega$ dispersion zones that allow for slow ($v_p<c$) or fast ($v_p>c$) phase propagation, where $k$ is the wavenumber, $\omega$ is angular frequency, $v_p$ is phase velocity, and $c$ is the speed of light in the medium of propagation. For example, smooth wavegudies (minimal periodic loading) are typical fast-wave systems, whereas some linear particle accelerators (linacs) are typical slow-wave systems, which allow the particles to travel synchronously with the accelerating phase crest of the rf field at $v_p {<} c$. 

Despite the rich variety of possible conditions of operation and geometric scalings, two scenarios, or extremes, are usually found in the literature that studies such structures. In one extreme, we find tranditional corrugated waveguides, which typically support a single (dominant) eigenmode, and whose period $b$ is typically shorter than the wavelength ($b<\lambda$). See Figure~\ref{fig:geometry} for the parameters $a,b,\delta$ and $r_0$ defining the geometry. Hybrid modes, which combine transverse electric (TE) and transverse magnetic (TM) modes, tend to dominate such structures and can have attractive propagation properties under the right conditions, \cite{ Borgnis,Mahmoud,jackson,Collin2,Clarricoats1,Clarricoats2}. Most literature treatements analyze this class of structures with a spectral expansion that assumes one dominant mode, with ohmic loss seen as the dominant mechanism of power loss. On the other extreme, at higher frequencies, the size of the geometric features of the periodic structure becomes much larger than $\lambda$, forcing the regime into multi-mode (also called overmoded) operation, where hundreds or thousands of modes could propagate simultaneously. In this limit, the structure becomes heavily diffractive (almost quasi-optical) and the dominant form of power loss, as seen from the perspective of the axial transport direction ($z$), is attributed to diffraction loss. Although Fresnel diffraction at knife-edges is an old subject in Fourier optics and scalar field theory, especially under plane-wave incidence assumptions, its effects become analytically and computationally more complicated in the presence of finite boundaries, such as vacuum chambers or cavity walls. Classical treatments, exemplified by the seminal work of J.~Schwinger and L.A.~Vainstein (see for example \cite{schwinger, schwingerWaveguides, SchwingerLevine, Vainstein1,Vainstein2,Vainstein3}), among others, which asymptotically analyze such periodic structures in the high-frequency limit, typically rely on methods such as factorization or Wiener-Hopf to find solutions, with the assumption that the periodic screens (irises) are infinitely thin (thickness $\delta$ taken as approximately zero). 

More realistic structures built out of conductive screens of nonzero thickness, $\delta$, will exhibit ohmic loss in the path of transportation, in addition to diffraction loss. To study this, scalar field theory (for diffraction and absorptive screens) must be superseded by vector field theory to include mode polarization effects (for ohmic losses and conductive screens). Therefore, the effect of the screen thickness $\delta$, especially when operating at wavelengths that are smaller or comparable in size to $\delta$, becomes a pressing concern. Recent investigations aimed at studying such effects were pursued at the SLAC National Accelerator Laboratory, Stanford University, and the European x-ray free-electron laser (European XFEL), Germany, where the transportation of THz radiation from a THz source over a long distance (e.g.~hundreds of meters) is required for pump-probe experiments \cite{DESY, NajiTHz1,NajiTHz2, Zhang,DESY3,DESY4}. The THz-pump pulse duration here is much shorter than the dimensions $a$ and $b$, and we are mainly interested in the transmission of the pulse itself (first pass), rather than any subsequent echoes that might follow \cite{NajiTHz1,NajiTHz2}. For a frequency in the range 3--15~THz, realistic implementations will have a screen thickness $\delta$ that is orders-of-magnitude larger than $\lambda$.  For example, at the lower wavelength of $0.1$~mm, where the diffraction loss of thin screens is highest, a structure with $b=33.3$~cm, $a=5.5$~cm, $\delta\leq 1$~mm, and $r_0=2a$ is predicted to have a total diffraction loss of $14\%$ over a 150-m distance (these dimensions were proposed for one structure candidate for THz transportation at SLAC) \cite{NajiTHz2, NajiTHz1, DESY}. Diffraction dominates the total loss, with ohmic loss on the screen edges (irises) being relatively negligible when we use such thin screens \cite{NajiTHz1,NajiTHz2}.  (Note that, for such oversized structures and short pulses, any part of the wave energy that is diffracted or scattered from the axial transportation direction into the shadow regions between the screens is considered lost. This allows one to model the conductive closed-chamber structure also as an open structure ($r_0\rightarrow \infty$), or a closed one with an absorptive chamber and screen side-walls \cite{NajiTHz1,NajiTHz2}.)

Conducting a formal vector field analysis that allows taking the structure into intermediate regimes of $\delta$, between the limit of $\delta\rightarrow0$ (Vainstein's benchmark) and the limit of $\delta\rightarrow b$ (smooth guide), however, exposes an interesting behavior that, to the best of our knowledge, has not been reported before. In this Letter we observe how ohmic loss can turn into the dominant loss mechanism, overtaking diffraction loss, when $\delta$ passes a certain value, even though the structure is still overmoded, oversized ($a,b,\delta\gg \lambda$), and still in the high frequency limit.

\begin{figure*}
\includegraphics[width=\textwidth]{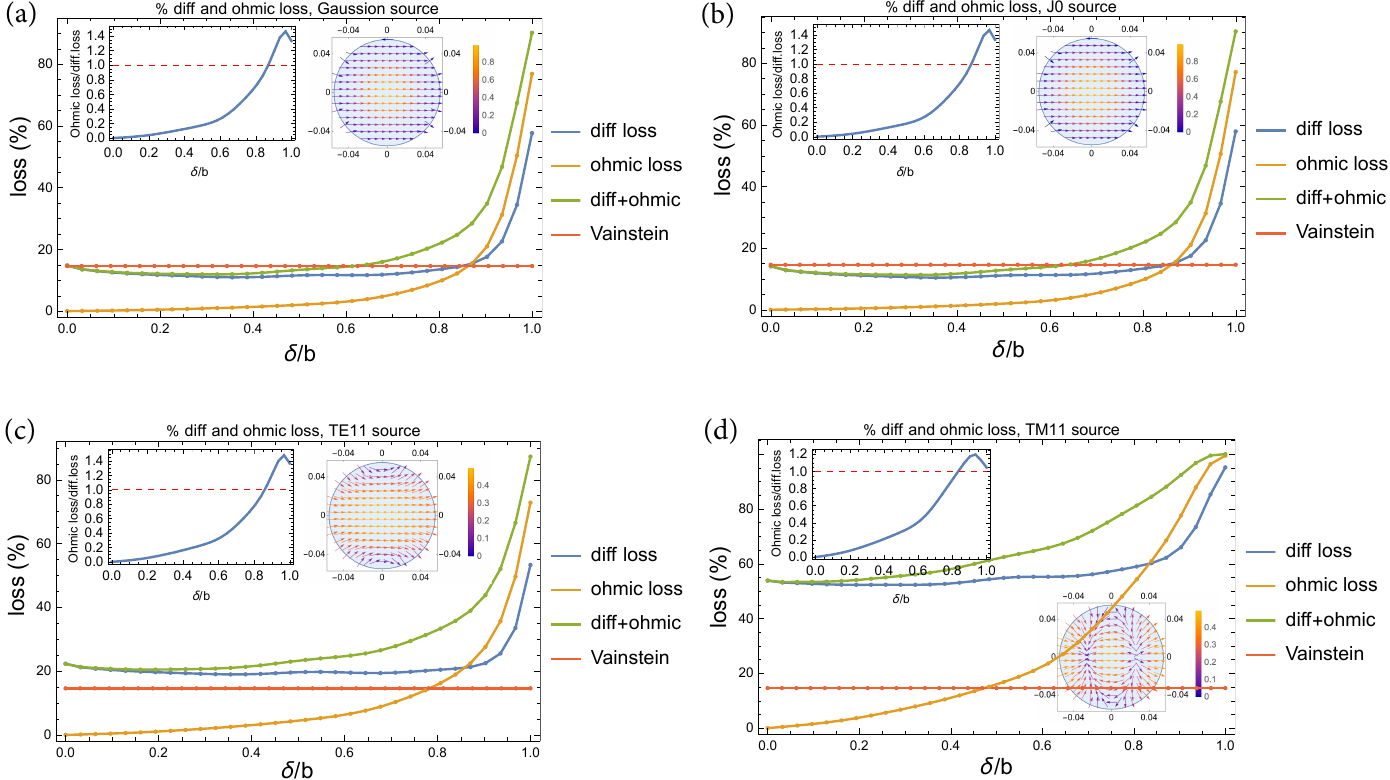}
\caption{\label{fig:allmodes}  Behavior of diffraction loss, ohmic loss, and combined loss as a function of $\delta$, for a structure with $\lambda=0.1$~mm, $b=33.3$~cm and $a=5.5$~cm, subject to 4 excitations: (a) Gaussian, (b) $J_0(2.4r/a)$ Bessel, (c) TE$_{11}$, and (d) TM$_{11}$ (see details in text). Each subfigure also contains a curve showing the ratio of ohmic loss to diffraction loss, and the mode's $E$ field plot.}
\end{figure*}

\emph{Field Analysis}---Although brute-force numerical simulation using standard tools (e.g.~finite element methods) for these overmoded structures are computationally prohibitive, solutions based on analytical spectral methods are possible and insightful. Field analysis for this structure, under finite values of $\delta$, has been conducted using two general approaches \cite{NajiTHz1,NajiTHz2}. The first solves for the eigensolutions of the boundary-value problem (steady state) by taking a long structure, with open annular shadow regions between the screens, and matching the mode expansions at the boundaries \cite{NajiTHz1}. This approach divides the structure radially into two regions, as shown in Figure~\ref{fig:geometry}(c). The second approach finds the scattering coefficients analytically at each discontinuity plane in the structure, then canscades the scattering effects \cite{NajiTHz2,NajiTHz3}. The space in this case is divided longitudinally into two regions (also called the cavity and waveguide regions), as shown in Figure~\ref{fig:geometry}(d). This approach has the advantage of revealing the transient regime at the entrance of the line and being relatively faster to implement computationally. Note that the oversized nature of the structure under paraxial incidence will imply that most of the scattering effects occur in the forward direction (forward-scatter) \cite{NajiTHz2,NajiTHz3}. For a harmonic time dependence $e^{-i\omega t}$ and the usual paraxial excitation, with dipolar polarization and slowly-varying envelope along $z$, we can show \cite{NajiTHz2,NajiTHz3} that the spectral (modal) expansion of the scattered fields are given analytically, for TE modes, by
\begin{align}
    E_{r,\text{TE}} &=\sum^{N}_{n=1}A_{n}\frac{\rho}{\nu'_{1n} r} J_{1}(\nu'_{1n}\frac{ r}{\rho})\cos\phi \ \zeta_\text{TE}(z)  \notag \\
   E_{\phi, \text{TE}} &=\sum^{N}_{n=1}-A_{n} J'_{1}(\nu'_{1n}\frac{ r}{\rho})\sin\phi \ \zeta_\text{TE}(z) \notag \notag\\ 
   E_{z,\text{TE}}&=0   \notag \\
   H_{r,\text{TE}}&=\sum^{N}_{n=1}\frac{- A_{n}}{Z_{0}}\left(\frac{\nu'^{2}_{1n}}{2k^{2}\rho^{2}}-1 \right)J'_{1}(\nu'_{1n}\frac{ r}{\rho})\sin\phi \ \zeta_\text{TE}(z) \notag \\
   H_{\phi, \text{TE}}&=\sum^{N}_{n=1}\frac{- A_{n} \rho}{Z_{0} r\nu'_{1n}}\left(\frac{\nu'^{2}_{1n}}{2k^{2}\rho^{2}}-1 \right)J_{1}(\nu'_{1n}\frac{ r}{\rho})\cos\phi \ \zeta_\text{TE}(z)   \notag 
   \end{align}
    \begin{align}
   H_{z,\text{TE}}&=\sum^{N}_{n=1}-A_{n}\frac{i\nu'_{1n}}{Z_{0}k \rho}J_{1}(\nu'_{1n}\frac{ r}{\rho})\sin\phi \ \zeta_\text{TE}(z) \notag 
\end{align}
where $i{\equiv}\sqrt{-1}$, $\zeta_\text{TE}(z){=}\exp (-iz\frac{\nu'^{2}_{1n}}{2k\rho^{2}}+ikz)$, $J_1$ is Bessel's function of the first kind and first order, $J'_1$ is its derivative with respect to its argument, $\nu_{1n}$ is the $n$th zero of $J_1$, $\nu'_{1n}$ is the $n$th zero of $J'_1$, $\rho$ is equal to either $a$ (in the waveguide region) or $r_0$ (in the cavity region), $\phi$ is the azimuthal angle, $Z_0$ is the wave impedance of free-space, and $N$ is the highest mode index in the expansion. Similarly, for TM modes we have
\begin{align}
    E_{r,\text{TM}}&=\sum^{N}_{n=1}-B_{n}J'_{1}\left( \nu_{1n}\frac{ r}{\rho}\right)\cos\phi \ \zeta_\text{TM}(z) \notag\\
    E_{\phi,\text{TM}}&=\sum^{N}_{n=1}\frac{B_{n} \rho}{\nu_{1n} r}J_{1}\left( \nu_{1n}\frac{ r}{\rho}\right)\sin\phi \ \zeta_\text{TM}(z) \notag \\
    E_{z,\text{TM}}&= \sum^{\infty}_{n=1}\frac{iB_{n}\nu_{1n}}{\rho k}J_{1}\left( \nu_{1n}\frac{ r}{\rho} \right)\cos\phi \ \zeta_\text{TM}(z)  \notag \\
    H_{r,\text{TM}}&=\sum^{\infty}_{n=1} \frac{-B_{n} \rho}{Z_{0}\nu_{1n} r}\left(\frac{\nu^{2}_{1n}}{2\rho^{2}k^{2}}+1 \right)J_{1}(\nu_{1n}\frac{ r}{\rho})\sin\phi \ \zeta_\text{TM}(z) \notag \\
    H_{\phi,\text{TM}}&=\sum^{N}_{n=1}\frac{-B_{n}}{Z_{0}}\left(\frac{\nu^{2}_{1n}}{2\rho^{2}k^{2}}+1\right)J'_{1}(\nu_{1n}\frac{ r}{\rho})\cos\phi \ \zeta_\text{TM}(z)  \notag \\
    H_{z,\text{TM}}&=0  \notag
\end{align}
where $\zeta_\text{TM}(z)=\exp(-iz\frac{\nu^{2}_{1n}}{2k\rho^{2}}+ikz)$. The scattering coefficients (denoted $A_n$ for the $n$th TE scattered mode, and $B_n$ for the $n$th TM scattered mode) on each type of discontinuity (``step-in" or ``step-out" boundaries, as the field travels to the right from region 1 to 2, then 1, and so forth, as shown in Figure~\ref{fig:geometry}(d)), assuming an incident TE or TM excitation of index $\ell$, are given by

\begin{align}
    A_n &=\begin{cases} 
        0, & \text{TM$_{\ell}$ on step-out}  \\
        \frac{-2B_{\ell} r_{0}J_{1}(\nu_{1\ell}a/r_{0})}{a \nu_{1\ell}(\nu'_{1n}-1/\nu'_{1n})J_{1}(\nu'_{1n})}, &  \text{TM$_{\ell}$ on step-in} \\
        \frac{2A_{\ell}\nu'_{1\ell} J_{1}(\nu'_{1\ell})J'_{1}(\nu'_{1n} a/r_{0})}{(\nu'^{2}_{1\ell}r_{0}^{2}/a^{2}-\nu'^{2}_{1n})(1-1/\nu'^{2}_{1n})J^{2}_{1}(\nu'_{1n})}, & \text{TE$_{\ell}$ on step-out} \\
        \frac{2A_{\ell}r_{0}^{2}\nu'_{1n} J'_{1}(\nu'_{1\ell} a/r_{0})}{a^{2}(\nu'^{2}_{1n}r_{0}^{2}/a^{2}-\nu'^{2}_{1\ell})(1-1/\nu'^{2}_{1n})J_{1}(\nu'_{1n})}, & \text{TE$_{\ell}$ on step-in} 
    \end{cases} \notag \\
\end{align}
and
\begin{align}
    B_n &=\begin{cases} \label{AnandBn}
        \frac{-2B_{\ell}a\nu_{1n}J_{0}(\nu_{1\ell})J_{1}\left( \nu_{1n} a/r_{0} \right)}{r_{0} (\nu_{1\ell}^2 r^2_0/ a^2-\nu_{1n}^2)J'^{2}_{1}\left( \nu_{1n}\right)}, & \text{TM$_{\ell}$ on step-out}  \\
        \frac{-2B_{\ell}a\nu_{1\ell}J_{0}(\nu_{1n})J_{1}(\nu_{1\ell}a/r_{0})}{r_{0}(\nu^{2}_{1n}-\nu^{2}_{1\ell}a^{2}/r_{0}^{2})J'^{2}_{1}(\nu_{1n})}, &  \text{TM$_{\ell}$ on step-in} \\
        \frac{-2A_{\ell}aJ_{0}(\nu'_{1\ell})J_{1}(\nu_{1n}a/r_{0})}{r_{0}\nu_{1n}J^{2}_{0}(\nu_{1n})}, & \text{TE$_{\ell}$ on step-out} \\
        0, & \text{TE$_{\ell}$ on step-in} 
    \end{cases}
\end{align}
\begin{figure}
\includegraphics[width=0.9\columnwidth]{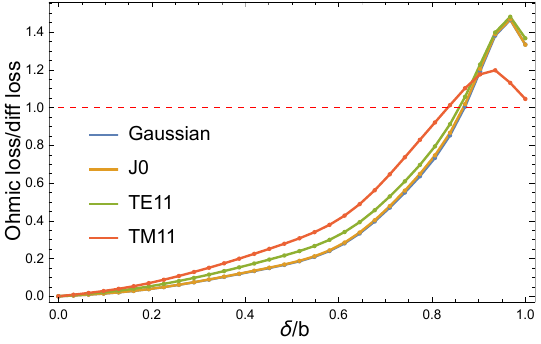}
\caption{\label{fig:ratio} A plot showing the evolution of the dominant loss character from diffractive to ohmic. The ratio of ohmic loss to diffraction loss is plotted as a function of $\delta$, for the 4 excitations. Note that the transition point is approximately fixed across the modes, despite their different transient regimes.}
\end{figure}

\emph{Observed Transition in Dominant Loss Mechanism}---Using these fields, we can now calculate the power lost to diffraction between the screens, as well as the power lost in the transportation path to ohmic heating at the edges of conductive screens \cite{NajiTHz2,NajiTHz3}, as a function of $\delta$. The evolution of diffraction and ohmic losses with $\delta$ is tracked and plotted in Figure~\ref{fig:allmodes} for four different types of excitations: a well-matched Bessel source, a nearly matched Gaussian source, a TE$_{11}$ source, and a TM$_{11}$ source. The Bessel source has a transverse $E$ field profile of $J_0(2.4 r/a)$ and is well-matched in reference to the eigenmode (steady state) in Vainstein's ideal limit, which has transverse $E$ field of the desirable form $J_{0}\left[\left(1-\varepsilon-i\varepsilon\right)\frac{2.4 r}{a}\right]$, where $\varepsilon\approx0.164\sqrt{\lambda b}/a$, \cite{DESY,Vainstein1}. This profile is akin to the $J_{0}(2.4r/a)$ profile, but with a small perturbation $\varepsilon$ that enters through its complex argument. The latter is a key feature of this geometry and represents the complex impedance equivalent to the structure's periodic nature, which enables the modes to mix and become hybrid in the axial region ($r\leq a$). The Gaussian dipole source has $1/e^2$ width that is backed-off to $\approx0.65 a$ to give a Gaussian profile of $\exp[-r^2/(0.65a)^2]$ that is better matched to the ideal $J_0(2.4 r/a)$ profile.

Figure~\ref{fig:ratio} summarizes the ratio of ohmic loss to diffraction loss for the 4 excitations. Even though the structure is still oversized and overmoded at the high-frequency limit, we observe the interesting transitional behavior, around the point with $\delta\approx 0.85 b$, where the ohmic loss overcomes diffraction loss to become the dominant mechanism of power loss on the line. This clear transition, which to the best of our knowledge has not been reported before, agrees with the expectation that the ohmic loss will become dominant for a smooth pipe in the limit of $\delta\rightarrow b$. The transition point $\delta\approx 0.85 b$ is equivalent to a shadow (cavity) region with an approximate width of the order of ${\sim}500\lambda$ (or ${\sim} 0.45a$), for a line that is 150-m long (the distance from source to the near experimental hall at SLAC's Linear Coherent Light Source, LCLS), with $a=5.5$~cm, $b=33.3$~cm, and $r_0\geq 2a$, at 3~THz. 

Interestingly, the transition point $\delta\approx 0.85 b$ seems to be approximately fixed, even when the exciting source fields are different and some of them (e.g.~the $J_0$ and Gaussian modes) have a much shorter transient regime, from the entrance of the line until they settle at the steady state, than the other modes (e.g.~TE$_{11}$ and TM$_{11}$). Before reaching the transition point, the ohmic loss increases steadily, while the diffraction loss is relatively plateaued; see Figure~\ref{fig:allmodes}. Beyond the transition point, it is seen that the ohmic loss increases sharply and becomes dominant, even though diffraction loss is also seen to independently increase (i.e.~calculated regardless of the presence of ohmic loss). 

This analysis is useful in bridging the two theoretical limits, showing exactly how a structure's dominant character transitions from diffractive to ohmic between these limits.

In addition to being theoretically interesting, this finding is important for experiments, as it reveals loss tradeoffs and informs experimental design. For instance, the physical realization with a Gaussian radiation source will have relatively relaxed mechanical requirements on how thin or thick the screens need to be, up to about $\delta\approx 0.6b$, when the total loss (ohmic and diffractive) starts to exceed the relatively steady ``plateau" around the Vainstein reference line; see Figure~\ref{fig:allmodes}(a). Conversely, the design of sections of the same waveguide structure with an added amount of attenuation can be achieved by simply choosing the appropriate $\delta (> 0.65b)$ for those sections, as shown in the same figure. 

Furthermore, the results highlight that, for well-matched excitation sources like $J_0(2.4r/a)$ or the backed-off Gaussian, the total loss (diffraction and ohmic combined) only becomes double the diffraction loss value at the ideal Vainstein limit ($\delta=0$) when $\delta$ is very thick ($\approx 0.9b$); see Figure~\ref{fig:allmodes}(a)-(b). Thus, as long as the irises are reasonably thin in practice (e.g.~$\delta<b/4$), one can consider mainly diffraction loss as the main loss, with its value approximately equal that at the the ideal limit of Vainstein ($\delta=0$), with ohmic loss then negligible.  Similar data can be derived for each cell or a line of any number of cells, using the same field equations listed above, and fully derived in \cite{NajiTHz2,NajiTHz3}.

\emph{Full Derivations and Data}---Full derivations and data related to this work can be found in \cite{NajiTHz3} for the full-scatter field analysis,  \cite{NajiTHz2} for the forward-scatter field analysis, and \cite{NajiTHz1} for the eigen-field analysis.

\emph{Acknowledgments}---The authors would like to thank Karl Bane, Sami Tantawi and Zhirong Huang, SLAC National Accelerator Laboratory, Stanford University, and Andrei Trebushinin, Eurpean XFEL, for discussions related to this work. Parts of this research were supported by Santa Clara University (GR103931-DENG1294Naji), while other parts by the US Department of Energy (contract number DE-AC02-76SF00515).

\hfill

\bibliography{myrefs}

\providecommand{\noopsort}[1]{}\providecommand{\singleletter}[1]{#1}%
\begin{thebibliography}{21}%
\makeatletter
\providecommand \@ifxundefined [1]{%
 \@ifx{#1\undefined}
}%
\providecommand \@ifnum [1]{%
 \ifnum #1\expandafter \@firstoftwo
 \else \expandafter \@secondoftwo
 \fi
}%
\providecommand \@ifx [1]{%
 \ifx #1\expandafter \@firstoftwo
 \else \expandafter \@secondoftwo
 \fi
}%
\providecommand \natexlab [1]{#1}%
\providecommand \enquote  [1]{``#1''}%
\providecommand \bibnamefont  [1]{#1}%
\providecommand \bibfnamefont [1]{#1}%
\providecommand \citenamefont [1]{#1}%
\providecommand \href@noop [0]{\@secondoftwo}%
\providecommand \href [0]{\begingroup \@sanitize@url \@href}%
\providecommand \@href[1]{\@@startlink{#1}\@@href}%
\providecommand \@@href[1]{\endgroup#1\@@endlink}%
\providecommand \@sanitize@url [0]{\catcode `\\12\catcode `\$12\catcode `\&12\catcode `\#12\catcode `\^12\catcode `\_12\catcode `\%12\relax}%
\providecommand \@@startlink[1]{}%
\providecommand \@@endlink[0]{}%
\providecommand \url  [0]{\begingroup\@sanitize@url \@url }%
\providecommand \@url [1]{\endgroup\@href {#1}{\urlprefix }}%
\providecommand \urlprefix  [0]{URL }%
\providecommand \Eprint [0]{\href }%
\providecommand \doibase [0]{https://doi.org/}%
\providecommand \selectlanguage [0]{\@gobble}%
\providecommand \bibinfo  [0]{\@secondoftwo}%
\providecommand \bibfield  [0]{\@secondoftwo}%
\providecommand \translation [1]{[#1]}%
\providecommand \BibitemOpen [0]{}%
\providecommand \bibitemStop [0]{}%
\providecommand \bibitemNoStop [0]{.\EOS\space}%
\providecommand \EOS [0]{\spacefactor3000\relax}%
\providecommand \BibitemShut  [1]{\csname bibitem#1\endcsname}%
\let\auto@bib@innerbib\@empty
\bibitem [{\citenamefont {Jackson}(1998)}]{jackson}%
  \BibitemOpen
  \bibfield  {author} {\bibinfo {author} {\bibfnamefont {J.~D.}\ \bibnamefont {Jackson}},\ }\href@noop {} {\emph {\bibinfo {title} {Classical Electrodynamics}}},\ \bibinfo {edition} {3rd}\ ed.\ (\bibinfo  {publisher} {Wiley},\ \bibinfo {year} {1998})\BibitemShut {NoStop}%
\bibitem [{\citenamefont {Collin}(1991)}]{Collin2}%
  \BibitemOpen
  \bibfield  {author} {\bibinfo {author} {\bibfnamefont {R.~E.}\ \bibnamefont {Collin}},\ }\href@noop {} {\emph {\bibinfo {title} {Field theory of guided waves}}},\ \bibinfo {edition} {2nd}\ ed.\ (\bibinfo  {publisher} {IEEE Press},\ \bibinfo {address} {New York},\ \bibinfo {year} {1991})\BibitemShut {NoStop}%
\bibitem [{\citenamefont {Chao}(1993)}]{Chao}%
  \BibitemOpen
  \bibfield  {author} {\bibinfo {author} {\bibfnamefont {A.}~\bibnamefont {Chao}},\ }\href@noop {} {\emph {\bibinfo {title} {Physics of Collective Beam Instabilities in High Energy Accelerators}}}\ (\bibinfo  {publisher} {Wiley},\ \bibinfo {year} {1993})\BibitemShut {NoStop}%
\bibitem [{\citenamefont {Slater}(1950)}]{slaterbook}%
  \BibitemOpen
  \bibfield  {author} {\bibinfo {author} {\bibfnamefont {J.~C.}\ \bibnamefont {Slater}},\ }\href@noop {} {\emph {\bibinfo {title} {Microwave Electronics}}}\ (\bibinfo  {publisher} {Dover Publishers},\ \bibinfo {year} {1950})\BibitemShut {NoStop}%
\bibitem [{\citenamefont {Milton}\ and\ \citenamefont {Schwinger}(2006)}]{schwinger}%
  \BibitemOpen
  \bibfield  {author} {\bibinfo {author} {\bibfnamefont {K.}~\bibnamefont {Milton}}\ and\ \bibinfo {author} {\bibfnamefont {J.}~\bibnamefont {Schwinger}},\ }\href@noop {} {\emph {\bibinfo {title} {Electromagnetic Radiation: Variational Methods, Waveguides and Accelerators}}}\ (\bibinfo  {publisher} {Springer},\ \bibinfo {year} {2006})\BibitemShut {NoStop}%
\bibitem [{\citenamefont {Mahmoud}(1991)}]{Mahmoud}%
  \BibitemOpen
  \bibfield  {author} {\bibinfo {author} {\bibfnamefont {S.~F.}\ \bibnamefont {Mahmoud}},\ }\href {https://digital-library.theiet.org/content/books/ew/pbew032e} {\emph {\bibinfo {title} {Electromagnetic Waveguides: theory and applications}}},\ Electromagnetic Waves\ (\bibinfo  {publisher} {Institution of Engineering and Technology},\ \bibinfo {year} {1991})\BibitemShut {NoStop}%
\bibitem [{\citenamefont {Naji}\ \emph {et~al.}(2022)\citenamefont {Naji}, \citenamefont {Stupakov}, \citenamefont {Huang},\ and\ \citenamefont {Bane}}]{NajiTHz1}%
  \BibitemOpen
  \bibfield  {author} {\bibinfo {author} {\bibfnamefont {A.}~\bibnamefont {Naji}}, \bibinfo {author} {\bibfnamefont {G.}~\bibnamefont {Stupakov}}, \bibinfo {author} {\bibfnamefont {Z.}~\bibnamefont {Huang}},\ and\ \bibinfo {author} {\bibfnamefont {K.}~\bibnamefont {Bane}},\ }\bibfield  {title} {\bibinfo {title} {Field analysis for a highly overmoded iris line with application to thz radiation transport},\ }\href {https://doi.org/10.1103/PhysRevAccelBeams.25.043501} {\bibfield  {journal} {\bibinfo  {journal} {Phys. Rev. Accel. Beams}\ }\textbf {\bibinfo {volume} {25}},\ \bibinfo {pages} {043501} (\bibinfo {year} {2022})}\BibitemShut {NoStop}%
\bibitem [{\citenamefont {Naji}\ and\ \citenamefont {Stupakov}(2022)}]{NajiTHz2}%
  \BibitemOpen
  \bibfield  {author} {\bibinfo {author} {\bibfnamefont {A.}~\bibnamefont {Naji}}\ and\ \bibinfo {author} {\bibfnamefont {G.}~\bibnamefont {Stupakov}},\ }\bibfield  {title} {\bibinfo {title} {Paraxial forward-scatter field analysis for a thz pulse traveling down a highly overmoded iris-line waveguide},\ }\href {https://doi.org/10.1103/PhysRevAccelBeams.25.123501} {\bibfield  {journal} {\bibinfo  {journal} {Phys. Rev. Accel. Beams}\ }\textbf {\bibinfo {volume} {25}},\ \bibinfo {pages} {123501} (\bibinfo {year} {2022})}\BibitemShut {NoStop}%
\bibitem [{\citenamefont {Naji}\ \emph {et~al.}(2025)\citenamefont {Naji}, \citenamefont {Gupta},\ and\ \citenamefont {Stupakov}}]{NajiTHz3}%
  \BibitemOpen
  \bibfield  {author} {\bibinfo {author} {\bibfnamefont {A.}~\bibnamefont {Naji}}, \bibinfo {author} {\bibfnamefont {P.~K.}\ \bibnamefont {Gupta}},\ and\ \bibinfo {author} {\bibfnamefont {G.}~\bibnamefont {Stupakov}},\ }\bibfield  {title} {\bibinfo {title} {Full-scatter vector field analysis of an overmoded and periodically-loaded \\ cylindrical structure for the transportation of thz radiation},\ }\href@noop {} {\bibfield  {journal} {\bibinfo  {journal} {ArXiv preprint: https://arxiv.org/abs/2506.09030}\ } (\bibinfo {year} {2025})}\BibitemShut {NoStop}%
\bibitem [{\citenamefont {Borgnis}\ and\ \citenamefont {Papas}(1958)}]{Borgnis}%
  \BibitemOpen
  \bibfield  {author} {\bibinfo {author} {\bibfnamefont {F.~E.}\ \bibnamefont {Borgnis}}\ and\ \bibinfo {author} {\bibfnamefont {C.~H.}\ \bibnamefont {Papas}},\ }\bibinfo {title} {Electromagnetic waveguides and resonators},\ in\ \href {https://doi.org/10.1007/978-3-642-45895-8_3} {\emph {\bibinfo {booktitle} {Elektrische Felder und Wellen}}}\ (\bibinfo  {publisher} {Springer Berlin Heidelberg},\ \bibinfo {address} {Berlin, Heidelberg},\ \bibinfo {year} {1958})\ pp.\ \bibinfo {pages} {285--422}\BibitemShut {NoStop}%
\bibitem [{\citenamefont {Clarricoats}\ \emph {et~al.}(1975{\natexlab{a}})\citenamefont {Clarricoats}, \citenamefont {Olver},\ and\ \citenamefont {Chong}}]{Clarricoats1}%
  \BibitemOpen
  \bibfield  {author} {\bibinfo {author} {\bibfnamefont {P.}~\bibnamefont {Clarricoats}}, \bibinfo {author} {\bibfnamefont {A.}~\bibnamefont {Olver}},\ and\ \bibinfo {author} {\bibfnamefont {S.}~\bibnamefont {Chong}},\ }\bibfield  {title} {\bibinfo {title} {Attenuation in corrugated circular waveguides. part 1: Theory},\ }\href@noop {} {\bibfield  {journal} {\bibinfo  {journal} {Proceedings of the Institution of Electrical Engineers}\ }\textbf {\bibinfo {volume} {122}},\ \bibinfo {pages} {1173} (\bibinfo {year} {1975}{\natexlab{a}})}\BibitemShut {NoStop}%
\bibitem [{\citenamefont {Clarricoats}\ \emph {et~al.}(1975{\natexlab{b}})\citenamefont {Clarricoats}, \citenamefont {Olver},\ and\ \citenamefont {Chong}}]{Clarricoats2}%
  \BibitemOpen
  \bibfield  {author} {\bibinfo {author} {\bibfnamefont {P.}~\bibnamefont {Clarricoats}}, \bibinfo {author} {\bibfnamefont {A.}~\bibnamefont {Olver}},\ and\ \bibinfo {author} {\bibfnamefont {S.}~\bibnamefont {Chong}},\ }\bibfield  {title} {\bibinfo {title} {Attenuation in corrugated circular waveguides. part 2: Experiment},\ }\href@noop {} {\bibfield  {journal} {\bibinfo  {journal} {Proceedings of the Institution of Electrical Engineers}\ }\textbf {\bibinfo {volume} {122}},\ \bibinfo {pages} {1180} (\bibinfo {year} {1975}{\natexlab{b}})}\BibitemShut {NoStop}%
\bibitem [{\citenamefont {Schwinger}\ and\ \citenamefont {Saxon}(1968)}]{schwingerWaveguides}%
  \BibitemOpen
  \bibfield  {author} {\bibinfo {author} {\bibfnamefont {J.}~\bibnamefont {Schwinger}}\ and\ \bibinfo {author} {\bibfnamefont {D.}~\bibnamefont {Saxon}},\ }\href@noop {} {\emph {\bibinfo {title} {Discontinuities in Waveguides: Notes on Lectures by Julian Schwinger}}},\ Documents on modern physics\ (\bibinfo  {publisher} {Gordon and Breach},\ \bibinfo {year} {1968})\BibitemShut {NoStop}%
\bibitem [{\citenamefont {Levine}\ and\ \citenamefont {Schwinger}(1948)}]{SchwingerLevine}%
  \BibitemOpen
  \bibfield  {author} {\bibinfo {author} {\bibfnamefont {H.}~\bibnamefont {Levine}}\ and\ \bibinfo {author} {\bibfnamefont {J.}~\bibnamefont {Schwinger}},\ }\bibfield  {title} {\bibinfo {title} {On the theory of diffraction by an aperture in an infinite plane screen. i},\ }\href {https://doi.org/10.1103/PhysRev.74.958} {\bibfield  {journal} {\bibinfo  {journal} {Phys. Rev.}\ }\textbf {\bibinfo {volume} {74}},\ \bibinfo {pages} {958} (\bibinfo {year} {1948})}\BibitemShut {NoStop}%
\bibitem [{\citenamefont {Vainstein}(1969)}]{Vainstein1}%
  \BibitemOpen
  \bibfield  {author} {\bibinfo {author} {\bibfnamefont {L.~A.}\ \bibnamefont {Vainstein}},\ }\href@noop {} {\emph {\bibinfo {title} {Open Resonators and Open Waveguides}}}\ (\bibinfo  {publisher} {Colem Press},\ \bibinfo {year} {1969})\BibitemShut {NoStop}%
\bibitem [{\citenamefont {Vainstein}()}]{Vainstein2}%
  \BibitemOpen
  \bibfield  {author} {\bibinfo {author} {\bibfnamefont {L.~A.}\ \bibnamefont {Vainstein}},\ }\bibfield  {title} {\bibinfo {title} {Open resonators for lasers},\ }\href@noop {} {\bibfield  {journal} {\bibinfo  {journal} {Sov. Phys. JETP}\ }\textbf {\bibinfo {volume} {17}},\ \bibinfo {pages} {709}}\BibitemShut {NoStop}%
\bibitem [{\citenamefont {Vainshtein}(1988)}]{Vainstein3}%
  \BibitemOpen
  \bibfield  {author} {\bibinfo {author} {\bibfnamefont {L.}~\bibnamefont {Vainshtein}},\ }\href@noop {} {\emph {\bibinfo {title} {Electromagnetic Waves}}}\ (\bibinfo  {publisher} {Radio i svyaz'},\ \bibinfo {address} {Moscow},\ \bibinfo {year} {1988})\ \bibinfo {note} {in Russian}\BibitemShut {NoStop}%
\bibitem [{\citenamefont {Geloni}\ \emph {et~al.}(2011)\citenamefont {Geloni}, \citenamefont {Kocharyan},\ and\ \citenamefont {Saldin}}]{DESY}%
  \BibitemOpen
  \bibfield  {author} {\bibinfo {author} {\bibfnamefont {G.}~\bibnamefont {Geloni}}, \bibinfo {author} {\bibfnamefont {V.}~\bibnamefont {Kocharyan}},\ and\ \bibinfo {author} {\bibfnamefont {E.}~\bibnamefont {Saldin}},\ }\href@noop {} {\bibinfo {title} {Scheme for generating and transporting thz radiation to the x-ray experimental floor at the lcls baseline}} (\bibinfo {year} {2011}),\ \Eprint {https://arxiv.org/abs/1108.1085} {arXiv:1108.1085 [physics.acc-ph]} \BibitemShut {NoStop}%
\bibitem [{\citenamefont {Zhang}\ \emph {et~al.}(2020)\citenamefont {Zhang}, \citenamefont {Fisher}, \citenamefont {Hoffmann}, \citenamefont {Jacobson}, \citenamefont {Kirchmann}, \citenamefont {Lee}, \citenamefont {Lindenberg}, \citenamefont {Marinelli}, \citenamefont {Nanni}, \citenamefont {Schoenlein}, \citenamefont {Qian}, \citenamefont {Sasaki}, \citenamefont {Xu},\ and\ \citenamefont {Huang}}]{Zhang}%
  \BibitemOpen
  \bibfield  {author} {\bibinfo {author} {\bibfnamefont {Z.}~\bibnamefont {Zhang}}, \bibinfo {author} {\bibfnamefont {A.~S.}\ \bibnamefont {Fisher}}, \bibinfo {author} {\bibfnamefont {M.~C.}\ \bibnamefont {Hoffmann}}, \bibinfo {author} {\bibfnamefont {B.}~\bibnamefont {Jacobson}}, \bibinfo {author} {\bibfnamefont {P.~S.}\ \bibnamefont {Kirchmann}}, \bibinfo {author} {\bibfnamefont {W.-S.}\ \bibnamefont {Lee}}, \bibinfo {author} {\bibfnamefont {A.}~\bibnamefont {Lindenberg}}, \bibinfo {author} {\bibfnamefont {A.}~\bibnamefont {Marinelli}}, \bibinfo {author} {\bibfnamefont {E.}~\bibnamefont {Nanni}}, \bibinfo {author} {\bibfnamefont {R.}~\bibnamefont {Schoenlein}}, \bibinfo {author} {\bibfnamefont {M.}~\bibnamefont {Qian}}, \bibinfo {author} {\bibfnamefont {S.}~\bibnamefont {Sasaki}}, \bibinfo {author} {\bibfnamefont {J.}~\bibnamefont {Xu}},\ and\ \bibinfo {author} {\bibfnamefont {Z.}~\bibnamefont {Huang}},\ }\bibfield  {title} {\bibinfo {title} {{A high-power, high-repetition-rate THz source for pump{--}probe
  experiments at Linac Coherent Light Source II}},\ }\href {https://doi.org/10.1107/S1600577520005147} {\bibfield  {journal} {\bibinfo  {journal} {Journal of Synchrotron Radiation}\ }\textbf {\bibinfo {volume} {27}},\ \bibinfo {pages} {890} (\bibinfo {year} {2020})}\BibitemShut {NoStop}%
\bibitem [{\citenamefont {Peetermans}\ \emph {et~al.}(2024)\citenamefont {Peetermans}, \citenamefont {Lemery}, \citenamefont {Hillert}, \citenamefont {Decking}, \citenamefont {Floettmann}, \citenamefont {Wernsmann}, \citenamefont {Trebushinin}, \citenamefont {Golubeva}, \citenamefont {Lockmann}, \citenamefont {Steffen}, \citenamefont {Czwalinna}, \citenamefont {Dohlus}, \citenamefont {Zagorodnov}, \citenamefont {Richards}, \citenamefont {Giesteira}, \citenamefont {Wohlenberg}, \citenamefont {Müller}, \citenamefont {Kalendar}, \citenamefont {Thoden}, \citenamefont {Müller}, \citenamefont {Wichmann}, \citenamefont {Walker}, \citenamefont {Guetg}, \citenamefont {Liu}, \citenamefont {Tomin}, \citenamefont {Qin}, \citenamefont {Long}, \citenamefont {Bielawski}, \citenamefont {Krasilnikov}, \citenamefont {Li}, \citenamefont {Geloni}, \citenamefont {Serkez}, \citenamefont {Lipka}, \citenamefont {Novokshonov}, \citenamefont {Kube}, \citenamefont {Hartl}, \citenamefont {Negodin},\ and\ \citenamefont
  {Scholz}}]{DESY3}%
  \BibitemOpen
  \bibfield  {author} {\bibinfo {author} {\bibfnamefont {K.}~\bibnamefont {Peetermans}}, \bibinfo {author} {\bibfnamefont {F.}~\bibnamefont {Lemery}}, \bibinfo {author} {\bibfnamefont {W.}~\bibnamefont {Hillert}}, \bibinfo {author} {\bibfnamefont {W.}~\bibnamefont {Decking}}, \bibinfo {author} {\bibfnamefont {K.}~\bibnamefont {Floettmann}}, \bibinfo {author} {\bibfnamefont {J.}~\bibnamefont {Wernsmann}}, \bibinfo {author} {\bibfnamefont {A.}~\bibnamefont {Trebushinin}}, \bibinfo {author} {\bibfnamefont {N.}~\bibnamefont {Golubeva}}, \bibinfo {author} {\bibfnamefont {N.}~\bibnamefont {Lockmann}}, \bibinfo {author} {\bibfnamefont {B.}~\bibnamefont {Steffen}}, \bibinfo {author} {\bibfnamefont {M.~K.}\ \bibnamefont {Czwalinna}}, \bibinfo {author} {\bibfnamefont {M.}~\bibnamefont {Dohlus}}, \bibinfo {author} {\bibfnamefont {I.}~\bibnamefont {Zagorodnov}}, \bibinfo {author} {\bibfnamefont {J.}~\bibnamefont {Richards}}, \bibinfo {author} {\bibfnamefont {F.}~\bibnamefont {Giesteira}}, \bibinfo {author} {\bibfnamefont
  {T.}~\bibnamefont {Wohlenberg}}, \bibinfo {author} {\bibfnamefont {L.~L.~G.}\ \bibnamefont {Müller}}, \bibinfo {author} {\bibfnamefont {V.}~\bibnamefont {Kalendar}}, \bibinfo {author} {\bibfnamefont {D.}~\bibnamefont {Thoden}}, \bibinfo {author} {\bibfnamefont {L.}~\bibnamefont {Müller}}, \bibinfo {author} {\bibfnamefont {R.}~\bibnamefont {Wichmann}}, \bibinfo {author} {\bibfnamefont {S.}~\bibnamefont {Walker}}, \bibinfo {author} {\bibfnamefont {M.}~\bibnamefont {Guetg}}, \bibinfo {author} {\bibfnamefont {S.}~\bibnamefont {Liu}}, \bibinfo {author} {\bibfnamefont {S.}~\bibnamefont {Tomin}}, \bibinfo {author} {\bibfnamefont {W.}~\bibnamefont {Qin}}, \bibinfo {author} {\bibfnamefont {T.}~\bibnamefont {Long}}, \bibinfo {author} {\bibfnamefont {S.}~\bibnamefont {Bielawski}}, \bibinfo {author} {\bibfnamefont {M.}~\bibnamefont {Krasilnikov}}, \bibinfo {author} {\bibfnamefont {X.}~\bibnamefont {Li}}, \bibinfo {author} {\bibfnamefont {G.}~\bibnamefont {Geloni}}, \bibinfo {author} {\bibfnamefont {S.}~\bibnamefont
  {Serkez}}, \bibinfo {author} {\bibfnamefont {D.}~\bibnamefont {Lipka}}, \bibinfo {author} {\bibfnamefont {A.}~\bibnamefont {Novokshonov}}, \bibinfo {author} {\bibfnamefont {G.}~\bibnamefont {Kube}}, \bibinfo {author} {\bibfnamefont {I.}~\bibnamefont {Hartl}}, \bibinfo {author} {\bibfnamefont {E.}~\bibnamefont {Negodin}},\ and\ \bibinfo {author} {\bibfnamefont {M.}~\bibnamefont {Scholz}},\ }\bibfield  {title} {\bibinfo {title} {{STERN}: accelerator-based {TH}z generation at {XFEL}}\ }(\bibinfo {organization} {Optical Terahertz Science and Technology 2024, Marburg (Germany), 8 Apr 2024 - 12 Apr 2024},\ \bibinfo {year} {2024})\BibitemShut {NoStop}%
\bibitem [{\citenamefont {Decking}\ \emph {et~al.}(2011)\citenamefont {Decking}, \citenamefont {Geloni}, \citenamefont {Kocharyan}, \citenamefont {Saldin},\ and\ \citenamefont {Zagorodnov}}]{DESY4}%
  \BibitemOpen
  \bibfield  {author} {\bibinfo {author} {\bibfnamefont {W.}~\bibnamefont {Decking}}, \bibinfo {author} {\bibfnamefont {G.}~\bibnamefont {Geloni}}, \bibinfo {author} {\bibfnamefont {V.}~\bibnamefont {Kocharyan}}, \bibinfo {author} {\bibfnamefont {E.}~\bibnamefont {Saldin}},\ and\ \bibinfo {author} {\bibfnamefont {I.}~\bibnamefont {Zagorodnov}},\ }\href {https://arxiv.org/abs/1112.3511} {\bibinfo {title} {Scheme for generating and transporting thz radiation to the x-ray experimental hall at the european xfel}} (\bibinfo {year} {2011}),\ \Eprint {https://arxiv.org/abs/1112.3511} {arXiv:1112.3511 [physics.acc-ph]} \BibitemShut {NoStop}%
\end{thebibliography}%

\end{document}